\documentclass[12pt]{iopart}
\usepackage{graphicx}
\usepackage{amsmath}
\usepackage{xcolor}
\begin{document}

\title[Magnetic interactions of Ni--Nb]{Absence of magnetic interactions in Ni–Nb ferromagnet–superconductor bilayers}

\author{Nathan Satchell$^{1,2}$, P. Quarterman$^3$, J.~A.~Borchers$^3$, Gavin Burnell$^2$, Norman O. Birge$^1$}

\address{$^1$Department of Physics and Astronomy, Michigan State University, East Lansing, Michigan 48912, USA}
\address{$^2$School of Physics and Astronomy, University of Leeds, Leeds LS2 9JT, United Kingdom}
\address{$^3$NIST Center for Neutron Research, National Institute of Standards and Technology, Gaithersburg, MD 20899, USA}
\ead{\mailto{N.D.Satchell@leeds.ac.uk}, \mailto{patrick.quarterman@gmail.com}}

\vspace{10pt}
\begin{indented}
\item[]February 2023
\end{indented}

\begin{abstract}
 Studies of ferromagnet-superconductor hybrid systems have uncovered magnetic interactions between the competing electronic orderings. The electromagnetic (EM) proximity effect predicts the formation of a spontaneous vector potential inside a superconductor placed in proximity to a ferromagnet. In this work, we use a Nb superconducting layer and Ni ferromagnetic layer to test for such magnetic interactions. We use the complementary, but independent, techniques of polarised neutron reflectometry and detection Josephson junctions to probe the magnetic response inside the superconducting layer at close to zero applied field. In this condition, Meissner screening is negligible, so our measurements examine only additional magnetic and screening contributions from proximity effects. We report the absence of any signals originating from EM proximity effect in zero applied field. Our observations indicate that either EM proximity effect is below the detection resolution of both of our experiments or may indicate a new phenomenon that requires extension of current theory. From our measurements, we estimate a limit of the size of the zero field EM proximity effect in our Ni-Nb samples to be $\pm0.27$~mT.

\end{abstract}

%
%
%
\maketitle
%
%

\section{\label{Intro}Introduction}

In the superconducting state, magnetic flux is expelled by the Meissner effect \cite{Meissner}. Hybrid ferromagnet-superconductor (\textit{F}-\textit{S}) systems can exhibit more complex screening properties due to the new physics present at the \textit{F}-\textit{S} interface \cite{flokstra2016remotely}. Recently, experimental observations via low energy muon spectroscopy reveal an otherwise unexpected contribution to flux expulsion in \textit{F}-\textit{S} bilayers and multilayers below the critical temperature ($T_{\text{c}}$) of the superconductor \cite{flokstra2016remotely, AnomalousMeissner, PhysRevB.100.020505, doi:10.1063/1.5114689, PhysRevB.104.L060506}. In those works Co and Nb are chosen for the \textit{F} and \textit{S} layers respectively. The experimental signature of this effect is an additional screening component originating at the \textit{F}-\textit{S} interface which increases the total flux screening inside the superconductor.

The development of the electromagnetic (EM) proximity effect theory by Mironov \textit{et al.} provides a framework in which many of the experimental observations can be interpreted \cite{ElectromagneticProximity}. As a result of the transparency of the \textit{F}-\textit{S} interface, electrons forming Cooper pairs can proximitise the \textit{F} layer. This proximity effect results in many well established transport phenomena, such as $\pi$ Josephson junctions \cite{RevModPhys.77.935}. Mironov \textit{et al.} showed that an added consequence of the proximity effect is that supercurrents flow inside the \textit{F} layer, giving rise to compensating Meissner supercurrents on the superconductor side of the interface. The observable result is a new component of screening at the \textit{F}-\textit{S} interface due to the presence of an additional vector potential. 

In the EM proximity effect theory, the additional internal field, $B_x$, inside the superconductor, at distance $x$ from the \textit{F}-\textit{S} interface, is described as \cite{ElectromagneticProximity},
\begin{equation}
\label{EM}
B_x = A_\text{EM} e^{-x/\lambda_\text{L}},
\end{equation}
\noindent where $A_\text{EM}$ is the strength of the EM proximity effect (proportional to the magnetisation of the \textit{F} layer) and $\lambda_\text{L}$ is the London penetration depth. The key observable predictions of the EM proximity effect are (i) $B_x$ decays with $\lambda_\text{L}$, (ii) $A_\text{EM}$ oscillates in amplitude and sign with the thickness of the \textit{F} layer, and (iii) $A_\text{EM}$ (and hence observable $B_x$) is present even in the absence of an applied magnetic field. The theoretical description of the additional screening currents caused by EM proximity effect are also considered and expanded to structures other than a \textit{F}-\textit{S} bilayer \cite{PhysRevB.64.134506, PhysRevB.99.144506, PhysRevB.99.104519, PhysRevB.100.134513, bespalov2022electromagnetic, PhysRevB.105.064510}. For example, the EM proximity effect also predicts the response of the \textit{F}-\textit{F}-\textit{S} superconducting spin-valve structure \cite{PhysRevB.99.104519} studied experimentally\cite{flokstra2016remotely}. For a review see \cite{mironov2021electromagnetic}.

Quantitatively, the EM proximity effect theory is applied to model the recent experimental observations of Flokstra \textit{et al.} in Co-Nb-Cu trilayers\cite{doi:10.1063/1.5114689}. It is found that Meissner screening alone cannot reproduce the experimental observation of a much enhanced screening in the trilayers compared to Nb and Nb-Cu samples. The depth dependent magnetic signal measured in the muon experiment is then modelled to include both traditional Meissner effect and the additional component of screening predicted by Equation \ref{EM}. The strength of the additional screening component in the Co-Nb-Cu system is found to be $A_\text{EM} = -0.9$~mT in a measurement field of 30~mT at 2.5~K.

Other proposed mechanisms for magnetic interactions in \textit{F}-\textit{S} systems include the induced ferromagnetism of Bergeret \textit{et al.} \cite{bergeret_induced_2004}. This theory considers a moment at the \textit{F}-\textit{S} interface caused by the local spatial distribution of Cooper pair spins. A number of reports attribute observations to this effect \cite{salikhov_experimental_2009, xia_inverse_2009, khaydukov_magnetic_2010, khaydukov_feasibility_2013}. The key observable of Bergeret \textit{et al.} is that the induced ferromagnetism decays over the coherence length ($\xi$), which for thin film Nb is much shorter than $\lambda_{L}$ \cite{PhysRevB.52.10395,PhysRevB.72.064503,PhysRevMaterials.4.074801}. Additionally, spin-triplet pairs can modify the magnetic response of \textit{F}-\textit{S} systems, for example, by introducing a paramagnetic Meissner component \cite{lofwander_interplay_2005,PhysRevB.70.052507, yokoyama_anomalous_2011, mironov_vanishing_2012, alidoust_meissner_2014,  PhysRevLett.116.127002}, observed experimentally by the muon technique \cite{di_bernardo_intrinsic_2015, rogers2021spin, PhysRevMaterials.5.114801}. 

Considerable research effort has also focused on what we describe here as stray field interactions, where the Meissner effect inside the superconductor acts to screen stray magnetic fields emerging from a ferromagnetic layer. Such stray fields are in some applications considered problematic, for example they can significantly distort the Fraunhofer pattern of ferromagnetic Josephson junctions reducing device functionality \cite{Bourgeois, PhysRevB.79.094523, PhysRevB.80.020506}. It is also possible, however, to engineer structures and devices in which the stray fields provide functionality, such as influencing the ferromagnetic domain structure or providing pinning sites for Abrikosov vortices \cite{tamegai_experimental_2011, doi:10.1063/1.4938467, del2015superconducting, PhysRevB.95.144504, curran2017continuously, marchiori2017reconfigurable, Pati_o_2018, satchell2019controlled, wang2018switchable}.

Additionally, there are several experimental observations of magnetic interactions in \textit{F}-\textit{S} systems where the underlying mechanism either falls outside of the categories outlined previously or remains unexplained \cite{PhysRevB.89.054510, PhysRevB.95.184509, PhysRevB.99.140503, PhysRevB.103.224521, 9625839, Nagy_2016}. For example, a previous report of Flokstra observed a measurable inverse proximity effect in a trilayer sample of Py-Nb-Py by the muon technique, but with a decay length from the \textit{F}-\textit{S} interface much shorter than either $\xi$ or $\lambda_{L}$ \cite{PhysRevB.89.054510}. 

In this work, we design and carry out two experiments to test a key prediction of the EM proximity effect: that the induced $B_x$ of Equation \ref{EM} inside the \textit{S} layer should be present in the absence of applied field. Nb is our superconductor of choice and Ni is chosen as the ferromagnet because (i) it has been extensively studied in Nb-Ni-Nb Josephson junctions \cite{PhysRevLett.89.187004,PhysRevB.73.174506,PhysRevLett.97.177003,PhysRevB.76.094522,4277679,PhysRevB.79.054501,baek2014hybrid,PhysRevApplied.7.064013,8359359,8666798}, (ii) the proximity in the Ni-Nb bilayer is expected to be in the clean (ballistic) limit \cite{PhysRevApplied.7.064013},  and (iii) the magnetic switching of thin Ni layers is relatively hard compared to the other elemental ferromagnets, such as Co (a requirement of our experiments).

The thickness of the Ni layer is fixed in both experiments at 2.8~nm. This thickness is guided by our preliminary measurements, see Supplementary Figure S1 \cite{SI}. In the first experiment, we use polarised neutron reflectometry (PNR) as a sensitive probe of screening and buried interfaces in our Ni-Nb bilayer. Previously, we have used PNR to measure $\lambda_{L}=96 \pm 9$~nm (uncertainty represents one standard error) inside a 200~nm thick Nb single layer film in the Meissner state \cite{PhysRevMaterials.4.074801}. However, unlike our previous study, here we add the Ni (2.8~nm) layer below the Nb (200~nm) layer. For this experiment we reduce the applied magnetic field as far as possible (while still retaining neutron polarisation) so that we are close to zero applied field. Minimising the applied field has the added benefit of removing contributions to the measured signal from the conventional Meissner effect, which keeps the data interpretation as simple as possible. Any changes in the PNR with the onset of the superconductivity can thus be attributed to magnetic interactions, such as the profile of Equation \ref{EM} expected from the EM proximity effect. 

Our second experiment uses detection Josephson junctions (DJJs), which are fabricated above Ni-Nb bilayers in direct electronic contact with the Nb following the geometry proposed by Mironov \textit{et al.} \cite{ElectromagneticProximity}. We set the Nb thickness here to 90~nm, so that we probe $\approx 1\lambda_{L}$ ($\approx 8\xi_\text{GL}$) from the \textit{F}-\textit{S} interface. We fabricate the DJJs with a Ru/Al multilayer barrier, which has a high interfacial resistance and strongly suppresses supercurrent. It is predicted that the EM proximity effect will be observable as a shift in the Fraunhofer pattern of the DJJ from zero applied field \cite{ElectromagneticProximity}. We compare the DJJ samples on the Ni layer to control samples where an insulating layer is placed between the Ni and Nb layers. 

We report the absence of any signals originating from EM proximity effect in near-zero applied field. Our observations indicate that either EM proximity effect is below the detection resolution of both of our experiments or may indicate a new phenomenon that requires extension of current theory. For each of our experimental techniques, we calculate upper limits on the size of $A_\text{EM}$.

\section{\label{Methods}Methods}

Thin films are deposited by sputtering with base pressure of $2 \times 10^{-8}$~Torr and partial water pressure of $3 \times 10^{-9}$~Torr ($4 \times 10^{-7}$~Pa), after liquid nitrogen cooling. We grow the films on 12.5 mm x 12.5 mm Si substrates, which have a typical native oxide layer. Growth is performed at an approximate Ar (6N purity) pressure of 2~mTorr and temperature of -25$^{\circ}$C. Triode sputtering is used for Nb, Ni, and Al from 57~mm diameter targets, and dc magnetron sputtering is used for Au and Ru from 24~mm diameter targets. The targets have 4N purity. Materials are deposited at typical growth rates of 0.4~nm~s$^{-1}$ for Nb and Au, and 0.2~nm~s$^{-1}$ for Ni, Al, and Ru. Growth rates are calibrated using an \textit{in situ} quartz crystal film thickness monitor and checked by fitting to Kiessig fringes obtained from x-ray reflectometry on reference samples. 

For PNR, we grow a series of Ni (2.8)-Nb (200) bilayer sheet films on Si substrates, where the nominal thicknesses are denoted in nanometers. The thickness of the Nb layer is fixed to that of our previous work so that the baseline screening properties are known \cite{PhysRevMaterials.4.074801}. 

For electrical transport samples, we fabricate standard ``sandwich" planer Josephson junctions using methodology described elsewhere \cite{PhysRevB.85.214522}. The full structure of the devices is Ni (2.8)-Nb (90)-[Ru (2)-Al (2)]$_6$-Ru (2)-Nb (5)-Au (5)-Nb (150). The Ni-Nb bilayer forms the bottom electrode of our device, where the Nb is chosen to be approximately one penetration depth ($1\lambda_\text{L}$) thick. The [Ru-Al]$_6$-Ru multilayer is the barrier for our DJJs and is chosen because the Ru/Al interfaces suppress supercurrent, allowing for tuning of the junctions' critical supercurrent ($I_c$). In this work, the junctions are circular with a designed diameter of 3 $\mu$m and the number of repeats is 6, which suppresses the critical current into the limit where the Josephson penetration depth is much larger than the junction diameter. The Nb (5)-Au (5) capping layers prevent oxidation during lithographic processing. The Nb (150) top electrode is deposited in the final stage of the processing. 

Further control samples are fabricated, where the Ni (2.8) layer was replaced by a Ni (2.8)-Nb (5)-Al$_2$O$_3$ (2.5) trilayer. The role of the Al$_2$O$_3$ insulator is to block any electronic proximity effects for control measurements and the thin non-superconducting Nb layer ensures consistency in the interfacial properties of the Ni layer.

We collect PNR using the Polarized Beam Reflectometer and Multi-Angle Grazing-Incidence K-vector reflectometer at the NIST Center for Neutron Research (NCNR). The incident and scattered neutron spins are polarised parallel or antiparallel to the applied in-plane magnetic field (\textit{H}) with supermirrors, and reflectivity is measured in the non-spin-flip cross sections ($R^{\uparrow \uparrow}$ and $R^{\downarrow \downarrow}$) as a function of the momentum transfer ($Q$) normal to the film surface. Given the incident beam is in the grazing configuration for the entire $Q$ range measured, the neutron beam effectively bathes the entire sample, and the data represent an ensemble average. The PNR data are reduced and modeled using the REDUCTUS \cite{reductus} software package and model-fit using the REFL1D program \cite{refl1d,refl1dweb}. The uncertainty of each fitting parameter is estimated using a Markov-chain Monte-Carlo simulation implemented by the DREAM algorithm in the BUMPS Python package\cite{DREAM}. Data are gathered at temperatures as low as 3 K, using a closed cycle refrigerator inserted into a 0.7 T electromagnet with field applied along the substrate orientation.

Electrical transport measurements are performed using a conventional four-point-probe measurement configuration with Keithley 6221 current source and 2182 nanovoltmeter \cite{NIST}. We collect transport data in a \textsuperscript{4}He cryostat with variable temperature insert (1.8 - 300~K) and 3~T superconducting split pair magnet. Magnetic characterisation is performed on a Quantum Design MPMS 3 magnetometer \cite{NIST} on cuttings of sister sheet film samples. 

\section{\label{Results}Results}

\subsection{\label{Characterisation}Ferromagnetic and superconducting properties of Ni-Nb bilayers}

\begin{figure}
        \includegraphics[width=0.49\textwidth]{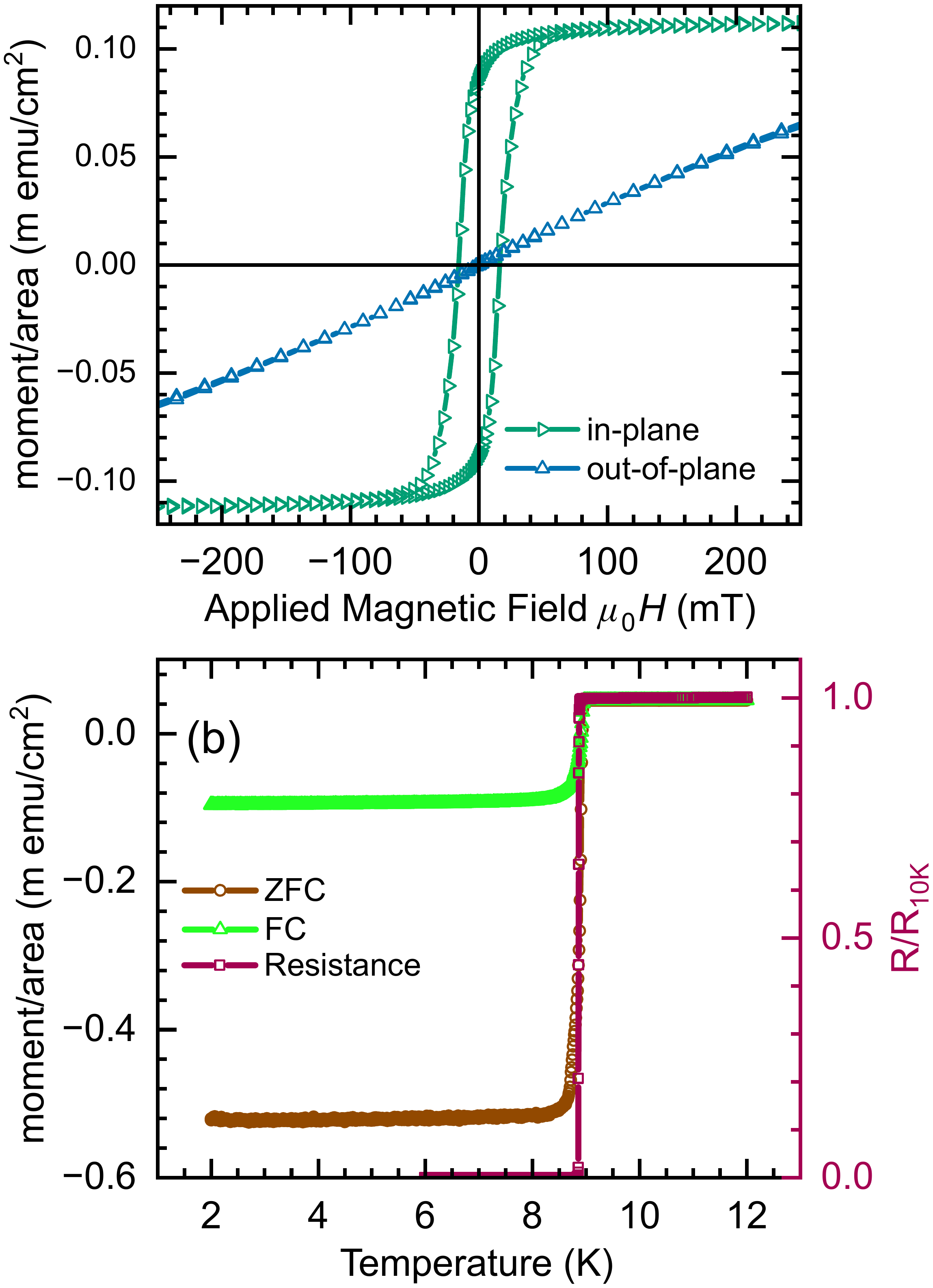}

        \caption{Magnetic and transport characterisation of the Si(sub)-Ni(2.8)-Nb(200) sheet film sample. (a) Magnetic hysteresis loops acquired at a temperature of 10 K with the applied field oriented in-plane and out-of-plane. The diamagnetic contribution from the substrate has been subtracted. Moment/area is calculated from the measured total moment and the measured area of the sample cutting. (b) Left axis: Moment versus temperature in an in-plane applied field of 1~mT in the zero field cooled (ZFC) and field cooled (FC) conditions. The temperature dependent signal is due to Nb screening the field. Right axis: Normalised resistance versus temperature. The $T_c$ of the sample from these measurements is taken either at the onset of the screening signal or the reduction in electrical resistivity to 50\% the normal state value and is $8.93\pm0.03$~K (uncertainty represents one standard error). Lines connecting the data points are guides.}
        \label{fig:mag}
\end{figure}

Figure \ref{fig:mag} shows measurements of the sheet film by magnetometry and transport to determine the magnetic switching behaviour and onset of superconductivity in our sample. Figure \ref{fig:mag} (a) shows the in- and out-of-plane magnetic hysteresis loops of the Si(sub)-Ni(2.8)-Nb(200) sample at 10~K. The diamagnetic signal from the substrate has been subtracted. The moment/area is calculated from the measured total moment of the sample and the measured area of the cutting used. The full field range of the acquired data are plotted in the Supplementary Information \cite{SI}. 

The in-plane easy axis loop and out-of-plane hard axis loop of Figure \ref{fig:mag} (a) indicates that the Ni has the expected in-plane magnetisation. For in-plane applied fields, the Ni layer has a remanence of $\approx75$\%, a coercive field of $\approx16$~mT, and a saturation field of $\approx200$~mT. The reduction of the remanence from 100\% and the large saturation field are suggestive that the Ni forms a multidomain structure when the applied field is removed. The volume magnetisation of the sample at saturation is calculated from the nominal thickness of the Ni layer to be $400\pm40$~emu/cm$^3$ (1~emu/cm$^3=1$~kA/m) (uncertainty represents one standard error), which is lower than the expected value of 485~emu/cm$^3$. We attribute the lower volume magnetisation of our sample to the formation of magnetic dead layers, which we estimate from the reduction in the measured volume magnetisation from the bulk value to have a thickness of 0.5~nm in our sample. The PNR results presented in Section \ref{PNR} and Table \ref{tab:PNRFit} confirm the presence of a magnetic dead layer in the sample at the SiO$_x$/Ni interface.

Figure \ref{fig:mag} (b) shows the moment versus temperature of the sample on the left axis obtained at 1~mT in-plane measurement field in the zero field cooled and field cooled conditions. At 1~mT in-plane field, the Ni is close to the remanent state and will be the major contribution to the signal above the $T_c$ of Nb. In the range of temperatures from 2-12~K the magnetisation of the Ni is not expected to change, so we attribute temperature dependent signal to the onset of superconductivity in the Nb layer. Plotted on the right axis is the normalised resistance of the sample. The $T_c$ of the sample from these measurements is taken either at the onset of the screening signal or the reduction in electrical resistivity to 50\% the normal state value and is $8.93\pm0.03$~K (uncertainty represents one standard error). This is a slight reduction from a comparable single Nb layer ($9.10\pm0.05$~K \cite{PhysRevMaterials.4.074801}). The full temperature range of the acquired transport data are plotted in the Supplementary Information \cite{SI}.

\subsection{\label{PNR}Polarised neutron reflectometry}

We perform PNR at as close to zero applied field as possible. In PNR, a small guide field is needed to maintain beam polarisation. The smallest possible applied field is found to be 1~mT, and is hence used as our measurement field. To limit concerns of flux trapping in the sample, we do not change field when below the transition temperature of the Nb (9 K). When changing field states, the temperature is increased to approximately 20 K. For reproducibility of the magnetic field condition, a saturating field of 700 mT is then applied, followed by lowering to the 1~mT guide field, and finally the sample is either measured at 20~K to provide the normal state condition or is cooled to the base temperature of 3~K in the superconducting state. 

\begin{figure}
        \includegraphics[width=0.75\textwidth]{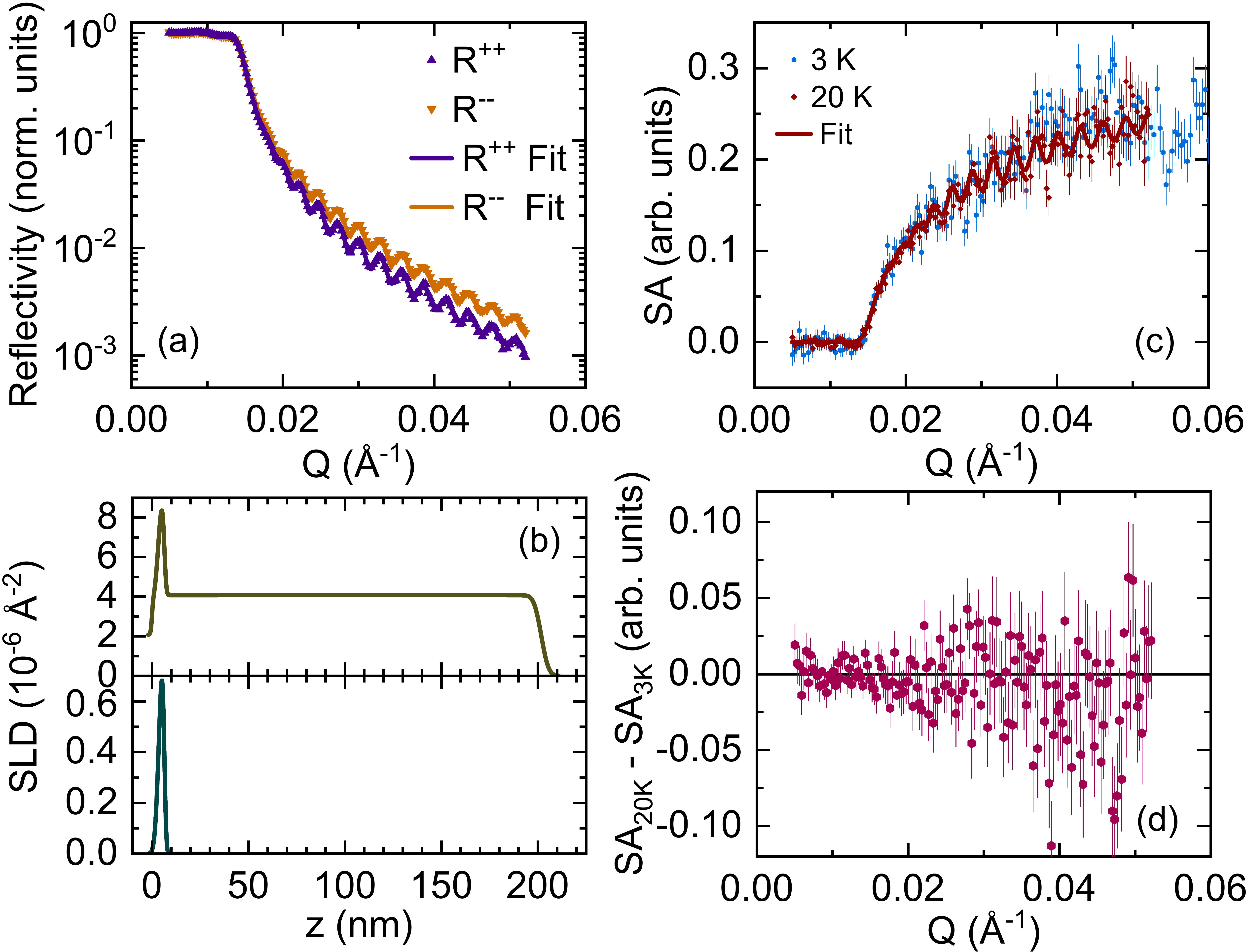}

        \caption{Polarised neutron reflectometry measurements of the Si(sub)-Ni(2.8)-Nb(200) sheet film sample at an applied field of 1~mT. (a) Non-spin-flip cross-section PNR data (points) with theoretical fits (line) at 20~K. (b) Nuclear (yellow, top) and magnetic (green, bottom) scattering length densities corresponding to the theoretical fits shown in (a). The best fit parameters are given in Table \ref{tab:PNRFit}. Z = 0 refers to the Si substrate surface. (c) The spin asymmetry at 20~K (red) and 3~K (blue). (d) The changes in the spin asymmetry between 20~K and 3~K. A horizontal line at $SA_{20K}-SA_{3K}=0$ is included to indicate to the reader that there are no significant changes to the PNR signal when the sample is cooled into the superconducting state. Presented uncertainties represent one standard error.}
        \label{fig:PNR1mT}
\end{figure}

Figure \ref{fig:PNR1mT} (a) shows the non-spin-flip cross-section PNR in the normal state (20~K) and the best fit to the reflectivity. At 20~K, the only expected magnetic contribution in the sample is the magnetisation of the Ni layer. To fully describe the reflectivity, a structural multilayer model is constructed, where each layer in the model has a thickness, roughness, and scattering length density fit parameter. Additionally, the magnetic scattering length density is fitted for the Ni layer. The best fit to the model is shown in Figure \ref{fig:PNR1mT} (a), where the structural and magnetic scattering length density depth profile corresponding to the model are given in Figure \ref{fig:PNR1mT} (b) and the best fit parameters in Table \ref{tab:PNRFit}. 

Modelling the 20~K data provides the following insights to our sample. The fitted layer thicknesses are close to the nominal growth thicknesses. At 20~K, the only magnetic contribution in the sample comes from the Ni layer with zero moment in the rest of the sample. The fitted magnetic dead layer thicknesses of 0.4~nm at the SiO$_x$/Ni interface and 0.0~nm at the Ni/Nb interface are consistent with the estimated dead layer thickness from magnetometry presented in Section \ref{Characterisation}.

Figure \ref{fig:PNR1mT} (c) and (d) show the main results of our PNR investigation into the EM proximity effect. The spin asymmetry [$SA = (R^{\uparrow \uparrow} - R^{\downarrow \downarrow})/(R^{\uparrow \uparrow} + R^{\downarrow \downarrow})$] for the Si(sub)-Ni(2.8)-Nb(200) sheet film sample at an applied field of 1~mT and temperatures of 20~K and 3~K are shown in Figure \ref{fig:PNR1mT} (c). In both the normal (20~K) and superconducting (3~K) states, there is zero spin asymmetry below the critical edge of the reflectivity, followed by an increasingly positive spin asymmetry with an oscillation. As expected, the fit corresponding to the structural and magnetic profile of Figure \ref{fig:PNR1mT} (b) describes the spin-asymmetry well. 

At 3~K, magnetic contributions from the superconducting state are expected in addition to that of the Ni layer. As we demonstrate in the Supplementary Information (Figure S2) \cite{SI}, at the 1~mT applied field, the Meissner contribution is too small to influence the PNR data in a measurable way. Therefore, when entering the superconducting state, only contributions from magnetic interactions, including the expected profile of Equation~\ref{EM}, influence the PNR. 

The key result of our PNR study is that comparing the measured spin asymmetry above and below the superconducting $T_c$, we observe no systematic changes with the onset of superconductivity. To be clear, the PNR data show no evidence of a magnetic proximity effect in near-zero field. That result is highlighted in Figure \ref{fig:PNR1mT} (d), where we plot the difference in spin asymmetry ($SA_{20K}-SA_{3K}$). The only deviations from $SA_{20K}-SA_{3K} =0$ are the scatter of individual data points. Such differences are not consistent with simulations of either a Meissner screening profile, Equation~\ref{EM}, or a combination of both. To confirm whether our data set is consistent with $SA_{20K}-SA_{3K} =0$, we perform a $\chi^2$ analysis relative to zero. The returned $\chi^2 = 175.15$ for 157 data points suggests that the difference in our data is indistinguishable from zero with $\approx 2\sigma$ confidence level. In a restricted region near the critical angle from $Q = 0.012$ to $0.022$ \AA$^{-1}$ where the difference should be most pronounced, the calculated $\chi^2 = 24.7$ for 34 data points, which gives a reduced $\chi^2$ smaller than 1.

\subsection{\label{JJs}Detection Josephson junctions}

\begin{figure}
        \includegraphics[width=0.49\textwidth]{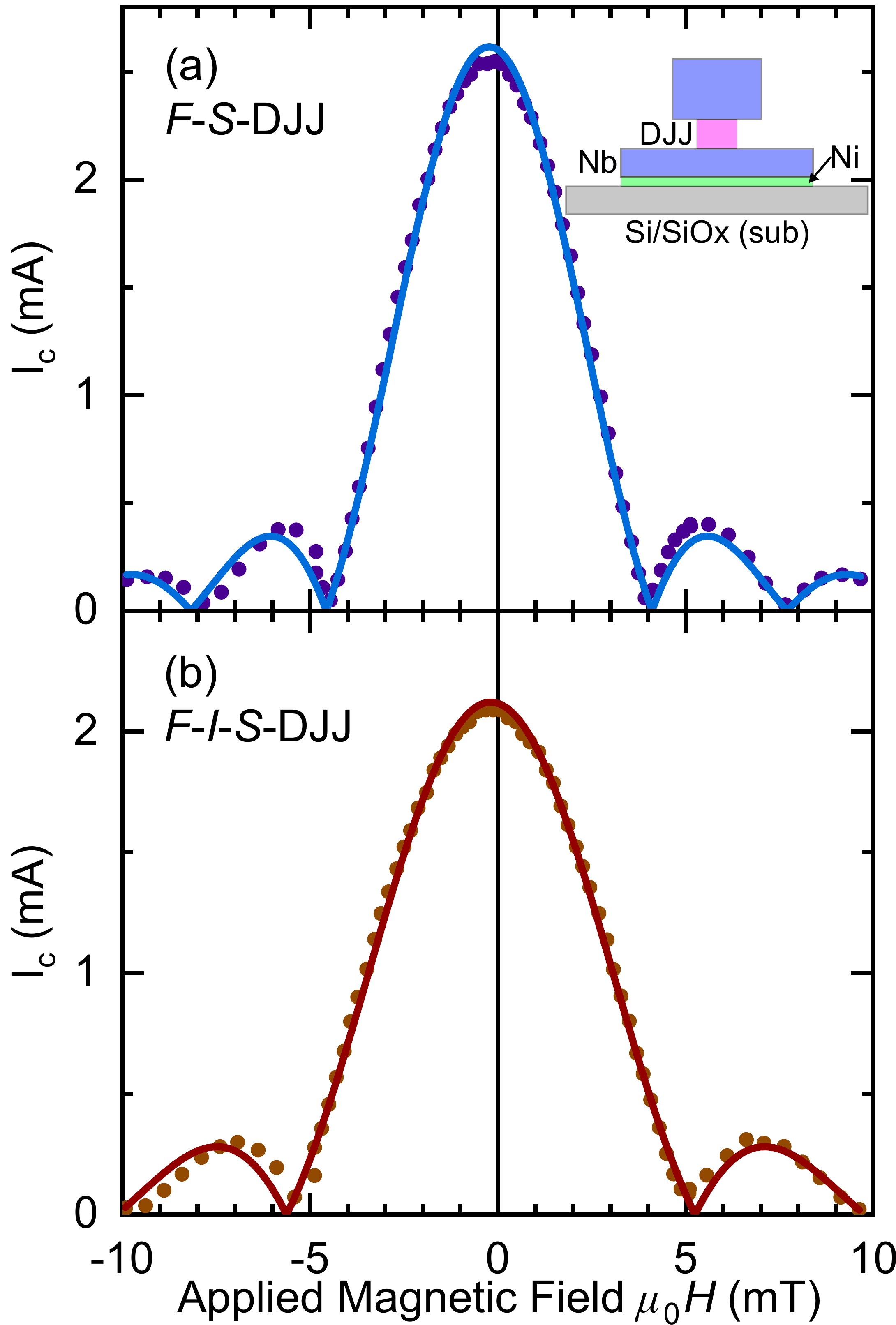}

        \caption{Representative $I_c(B)$ Fraunhofer patterns for the detection Josephson junction (DJJ) devices at 1.8~K. (a) The DJJ is placed on a Ni (2.8)-Nb (90) bilayer. (b) A control sample where an Al$_2$O$_3$ (2.5 nm) insulator layer is added between the Ni and Nb. The error in determining $I_c$ is smaller than the data points. Solid lines show fit to Equations \ref{Airy} and \ref{Phi} to extract $H_\text{shift}$, the amount the Fraunhofer pattern is shifted from zero applied field. Both devices show a small negative $H_\text{shift}$.}
        \label{fig:DJJ1}
\end{figure}

DJJs are fabricated above the Ni (2.8)-Nb (90) bilayer. The geometry of the devices is shown in the inset of Figure \ref{fig:DJJ1} (a) and our DJJs are designed to closely follow the devices theoretically proposed to measure the EM proximity effect in reference \cite{ElectromagneticProximity}. The design width of the bottom electrode, containing the Ni/Nb bilayer, is 18~$\mu$m. The length of the bottom electrode is 2400~$\mu$m and six DJJs are placed along the length of the bottom electrode with design widths of 3~$\mu$m. In-plane magnetic fields are applied along the long axis of the bottom electrode.  

Figure \ref{fig:DJJ1} shows representative $I_c(B)$ Fraunhofer response of our DJJs. The Fraunhofer patterns of two samples are presented. In the first sample, the superconducting Nb layer is in direct electronic contact with the Ni layer. In the second control sample, the Nb layer is separated from the Ni layer by a dielectric Al$_2$O$_3$ layer, which removes any electronic proximity effects. For circular Josephson junctions, the $I_c$(B) response can be described by the Airy function \cite{barone1982physics},
\begin{equation}
\label{Airy}
I_c = I_{c0} \left | 2J_1 (\pi \Phi / \Phi_0)/(\pi \Phi / \Phi_0) \right |,
\end{equation}
\noindent where $I_{c0}$ is the maximum critical current, $J_1$ is a Bessel function of the first kind, $\Phi_0=h/2e$ is the flux quantum, and $\Phi$ is the flux through the junction. In our case, 
\begin{equation}
\label{Phi}
\Phi = \mu_0 (H_\text{app} - H_\text{shift}) w (2\lambda_\text{L} + d),
\end{equation}
\noindent where $w$, $\lambda_\text{L}$ and $d$ are the width of the junction, the effective London penetration depth of the electrodes \cite{barone1982physics}, and the total thickness of all the normal metal layers and \textit{F} layers in the junction. $H_\text{app}$ is the applied field and $H_\text{shift}$ is the key parameter we report here, the amount $I_{c0}$ is shifted from $H$ = 0. We expect that $H_\text{shift}$ corresponds to the condition when the external applied field cancels the internal field from the EM proximity effect ($B_x$) at the junction, $\mu_0 H_\text{shift}=-B_x$.  

For consistency, prior to each measurement we follow the following initialisation routine. The samples are first held above the $T_c$ of Nb at 12~K and magnetised with an in-plane applied field of $+1$~T to fully saturate the Ni layer. The field is applied in the long axis of the device's bottom electrode. The saturation field is removed and we cool the sample in zero applied field (in practice there will inevitably be a small remanent field due to trapped flux in the magnet). Once the temperature reaches 1.8~K (the base temperature of our cryostat), the samples are measured by recording the $I-V$ characteristic to extract $I_c(B)$ at each applied field in the range 10 to -10~mT. After the measurement sequence is complete, the sample is warmed to 12~K and we repeat the cycle.

The $I_c(B)$ Fraunhofer measurements shown in Figure \ref{fig:DJJ1} are representative examples of $I_c(B)$ device behaviour. We performed multiple measurements on a total of 6 devices (3 \textit{F}-\textit{S} and 3 \textit{F}-\textit{I}-\textit{S} control samples) to calculate average values of $\overline{H}_\text{shift}$. For the \textit{F}-\textit{S} devices, $\mu_0\overline{H}_\text{shift}=-0.26\pm0.02$~mT. For the \textit{F}-\textit{I}-\textit{S} control devices, $\mu_0\overline{H}_\text{shift}=-0.23\pm0.03$~mT. Therefore, although we do observe that $\mu_0\overline{H}_\text{shift} \neq 0$, there is not an obvious signal attributable to the EM proximity effect as the $\mu_0\overline{H}_\text{shift}$ is observed in the control samples also.

There are two potential sources of the observed non-zero $\overline{H}_\text{shift}$ in our experiment which would be expected to be present for both the \textit{F}-\textit{S} and \textit{F}-\textit{I}-\textit{S} control samples. The first is trapped flux in the superconducting coils which apply the measurement field. To explore this contribution, we measure identical DJJ device without the Ni layer. Any $H_\text{shift}$ observed in the sample without the Ni layer is related to the trapped flux in our superconducting coils. As shown in Supplemental Figure S4, the observed $\mu_0 H_\text{shift} = 0.15$~mT indicates that a small positive applied field must be applied to cancel the trapped flux and achieve the zero field condition. Interestingly, the direction of $H_\text{shift}$ observed in the sample without the Ni layer is opposite to the direction of ${H}_\text{shift}$ observed in the devices with the Ni layer. 

A second possible source of non-zero $\overline{H}_\text{shift}$ are stray fields from the Ni layer. There are many possible sources of stray fields from the Ni layer including any combination of return fields, the edges of the devices, domain walls, and orange peel roughness \cite{schrag2000neel}. We believe that stray fields are the most likely source of our observed small negative $\overline{H}_\text{shift}$.

The differences in the $I_c(B)$ and width of the central lobe of the Fraunhofer patterns of the two presented devices shown in Figure \ref{fig:DJJ1} (a) and (b) is due to the sample to sample variation in defining the diameter of the DJJ by photolithography. The fitted diameter of the \textit{F}-\textit{S} sample is 2.9~$\mu$m and the \textit{F}-\textit{I}-\textit{S} control sample is 2.3~$\mu$m.

\section{Discussion}

\begin{figure}
        \includegraphics[width=0.49\textwidth]{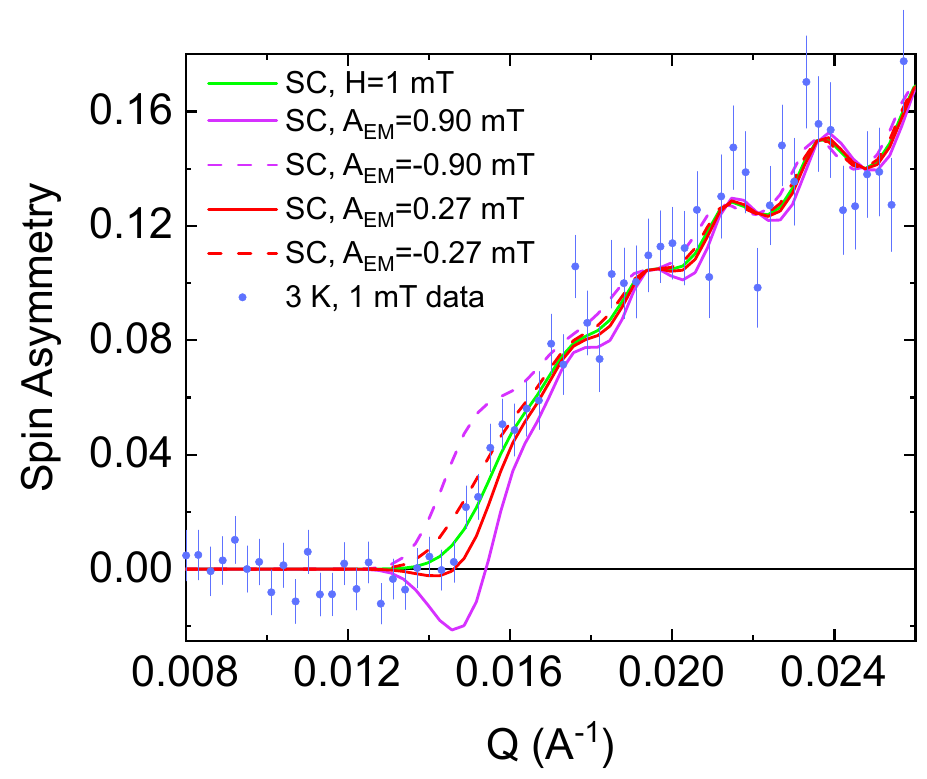}

        \caption{The spin asymmetry of the Si(sub)-Ni(2.8)-Nb(200) film at 3~K (color circles) with model fits assuming superconductivity with a London penetration depth of 96.2~nm in a 1~mT field. The effects of an additional proximity effect are considered for with $A_{EM} =$ $\pm$0.27~mT (solid and dashed color lines) as well as $\pm$0.9~mT (solid and dashed color lines). Presented uncertainties represent one standard error.}
        \label{fig:PNRProx}
\end{figure}

PNR simulations of Equation \ref{EM} applied to our Ni-Nb bilayer imply that changes in the sample magnetisation or excluded magnetic field due to the EM proximity effect should manifest in the spin asymmetry, as this quantity is sensitive to the magnetic induction in the sample. By comparing the fidelity of our experimental data in Figure \ref{fig:PNR1mT} to the results of these simulations, we do not see any evidence of additional magnetism from proximity effects but can estimate an approximate limit on the size of $A_\text{EM}$ at zero applied field. First we note that the field expulsion expected with a London penetration depth of 96.2 nm from the Meissner effect \cite{PhysRevMaterials.4.074801} at 3~K in a 1~mT field does not detectably alter the model for the reflectivity and spin asymmetry. (As shown in supplementary Figures S2 and S3, this behavior contrasts with that observed in a larger field of 150~mT \cite{SI}.) 
Next we consider the variation in the spin asymmetry that is expected if $A_{EM} = \pm 0.9$~mT, which matches the magnitude of the effect observed by Flokstra {\it et al} in Co (2.4)-Nb bilayers \cite{doi:10.1063/1.5114689}. As a reminder, $A_{EM}$ in Equation \ref{EM} is proportional to the Ni magnetisation in our case, which was determined from the PNR fit shown in Figure \ref{fig:PNR1mT} (a) and (c) to be 317 emu/cm$^3$ (Table \ref{tab:PNRFit}) in the 1~mT remanent field.  As seen in Figure \ref{fig:PNRProx}, the models for these values of $A_{EM}$ deviate substantially from the 3~K PNR spin asymmetry data, especially in the region near the critical angle. The magnitude of $A_{EM}$ was gradually decreased in order to approximate the sensitivity of the PNR measurement. Figure \ref{fig:PNRProx} shows the model fits calculated for $A_{EM}  = \pm$ 0.27~mT, which seem to be just above the detection limit for this technique. 

The Josephson junction experiment can also be used to estimate the limit of the EM proximity effect and the parameter $A_\text{EM}$. The Fraunhofer pattern of the DJJ is centered where the total field in the junction is zero. Assuming no stray fields or trapped flux, in the presence of the EM proximity effect the total field in the DJJ will be zero when the external applied field cancels the internal field of the EM proximity effect, which from Equations 1 and 3 is the condition $\mu_0H_\text{shift}=-B_x$. If $\mu_0H_\text{shift}=0$, we can estimate the limit of $A_\text{EM}$ from the experimental uncertainty in determining $\mu_0H_\text{shift}$.
In our experiment, due to the presence of stray fields and/or trapped flux, $\mu_0H_\text{shift}\neq0$ for either the \textit{F}-\textit{S} or \textit{F}-\textit{I}-\textit{S} control samples. In this case, it is the difference in $\mu_0\overline{H}_\text{shift}$ between the two sample geometries which provides $B_x$ and hence $A_\text{EM}$. The experimental uncertainty in determining $\mu_0\overline{H}_\text{shift}$ from the \textit{F}-\textit{S} DJJ devices is $\pm0.02$~mT and from the \textit{F}-\textit{I}-\textit{S} control samples is $\pm0.03$~mT. Therefore, it is reasonable to expect to be able to observe changes in $\mu_0\overline{H}_\text{shift}$ between the sample geometries that are of the order $\pm0.1$~mT. A difference of $\pm0.1$~mT would correspond to $A_\text{EM}$ at the Ni-Nb interface of $\pm0.27$~mT, which we conclude from both the DJJ and PNR techniques to be the upper limit of the EM proximity effect at zero applied field in our samples.

The absence of detectable $A_\text{EM}$ in the PNR and DJJ measurements suggests that any signals of the EM proximity effect at zero applied field are below the observable limit resolution of our presented measurements, which we estimate to be $A_\text{EM}<\pm0.27$~mT. By comparison to this limit, Flokstra \textit{et al.} report $A_\text{EM} = -0.9$~mT in Co (2.4)-Nb bilayers by the muon technique \cite{doi:10.1063/1.5114689}. An $A_\text{EM}$ of this magnitude, if present in our samples, should be observable in both of our experiments. 

Alternatively, the absence of detectable $A_\text{EM}$ in our experiments may indicate a new phenomenon that requires extension of current theory. An important difference between the experimental results we report in this work and the previous \textit{F}-\textit{S} experimental works using the muon technique is the coexistence of conventional screening currents in those previous works. In the muon technique, the measurements are performed in an applied magnetic field, typically 10-40~mT, and the reported observations of \textit{F}-\textit{S} proximity effect manifests as a modification of the conventional Meissner screening \cite{flokstra2016remotely, AnomalousMeissner, PhysRevB.100.020505, doi:10.1063/1.5114689, PhysRevB.104.L060506, di_bernardo_intrinsic_2015, rogers2021spin, PhysRevMaterials.5.114801}. Our observations may imply that the expected EM proximity effect manifests only when there are coexisting screening currents, and not in the zero field condition of this work. The zero field muon experiment has not been attempted, but the applied field dependence of the excess screening reported by Flokstra \textit{et al.} may support such conclusions \cite{AnomalousMeissner}. 

High field measurements using the DJJs is not possible due to the Fraunhofer physics of the junction. On the other hand, PNR can be performed at high field and exploratory measurements in an applied field of 150~mT are reported in Supplemental Figure S3 \cite{SI}. The PNR at 150~mT shows a pronounced difference upon entering the superconducting state for our sample with the Ni adjacent to the Nb layer and for a second sample with the Ni and Nb separated with an insulating spacer layer. The Ni-Nb data are not well described by a model accounting for only simple Meissner screening in the superconducting state. We present a fit to the Ni-Nb data with a linear combination of the Meissner and EM proximity formalisms in the superconducting state, which are not inconsistent with a finite $A_\text{EM}$. However, to fully describe the data a model accounting for vortices is required.

The observation in our DJJ experiments of $\overline{H}_\text{shift}$ in both our proximity samples and control samples (in which the proximity effect is suppressed by the addition of an insulating layer) highlights the importance of control experiments in this field.

\section{Conclusions}

In conclusion, we have performed polarised neutron reflectometry and detection Josephson junction measurements on a bilayer sample of Ni(2.8 nm)/Nb. Our measurements probe for additional contributions to the screening and magnetisation of the samples below the superconducting transition temperature at close to zero applied field. We report the absence of any signals originating from EM proximity effect in zero applied field. Our observations indicate that either EM proximity effect is below the detection resolution of both of our experiments or may indicate a new phenomenon that requires extension of current theory. From our measurements, we estimate a limit of the size of the zero field Electromagnetic Proximity Effect in our Ni-Nb samples to be $\pm0.27$~mT.  

The data associated with this paper are openly available from the NCNR and University of Leeds data repositories \cite{Data}.

\ack
We are grateful to Brian J. Kirby for helpful, in-depth discussions. We acknowledge the support of the EPSRC through Grant No. EP/V028138/1.
P.Q. acknowledges support from the National Research Council Research Associateship Program. This project has received partial funding from the European Unions Horizon 2020 research and innovation programme under the Marie Sk\l{}odowska-Curie Grant Agreement No. 743791 (SUPERSPIN).

\section*{References}
\bibliography{EMProx.bib}

\begin{thebibliography}{10}
\bibitem{NIST}
{Certain commercial products or company names are identified here to describe
  our study adequately. Such identification is not intended to imply
  recommendation or endorsement by the National Institute of Standards and
  Technology, nor is it intended to imply that the products or names identified
  are necessarily the best available for the purpose.}
\end{thebibliography}

\clearpage

\clearpage

\clearpage

\clearpage

\clearpage

\begin{table}[]
\begin{tabular}{|l|l|l|l|l|}
\hline
Layer           & Thickness                 & Roughness                 & Nuclear SLD                  & Magnetic SLD \\
                & \multicolumn{1}{c|}{nm} & \multicolumn{1}{c|}{nm} & \multicolumn{1}{c|}{10$^{-6}$ \AA$^{-2}$} & 10$^{-6}$ \AA$^{-2}$ (emu/cm$^3$)      \\ \hline
Nb              & $195.3\pm0.2$                     & $3.0\pm0.1$                       & $4.07\pm0.02$                        & 0            \\ \hline
Ni              & $2.7\pm0.2$                       & $0.8\pm0.2$                       & $9.5\pm0.1$                        & $0.9\pm0.1$ ($320\pm30$)       \\ \hline
Ni (dead layer) & $0.4\pm0.25$                       &                           & $9.5\pm0.1$                        & 0            \\ \hline
SiO$_x$            & $3.3\pm0.25$                       & $1.7\pm0.3$                       & $3.6\pm0.3$                        & 0            \\ \hline
Si              &                           & 0.5                       & 2.069                        & 0            \\ \hline
\end{tabular}
\caption{Best fit parameters corresponding to the 20~K PNR model shown in Figure~\ref{fig:PNR1mT}. The uncertainty of each fitting parameter is estimated using the DREAM algorithm, see text.}
\label{tab:PNRFit}
\end{table}

\clearpage

\end{document}


\begin{center}
\large{Supplementary information for: ``Absence of magnetic interactions in Ni–Nb
ferromagnet–superconductor bilayers''}\\
\small{Nathan Satchell$^{1,2}$, P. Quarterman$^3$, J.~A.~Borchers$^3$, Gavin Burnell$^2$, Norman O. Birge$^1$\\$^1$ Department of Physics and Astronomy, Michigan State University, East Lansing, Michigan 48912, USA  \\$^2$School of Physics and Astronomy, University of Leeds, Leeds, LS2 9JT, United Kingdom  \\$^3$NIST Center for Neutron Research, National Institute of Standards and Technology, Gaithersburg, MD 20899, USA}
\end{center}



This supplementary information contains additional analysis and data in support of the conclusions presented in the main text.

\section{Preliminary Josephson junction study of $\text{Ni}$ thickness dependence}

\begin{figure}
        \includegraphics[width=0.49\textwidth]{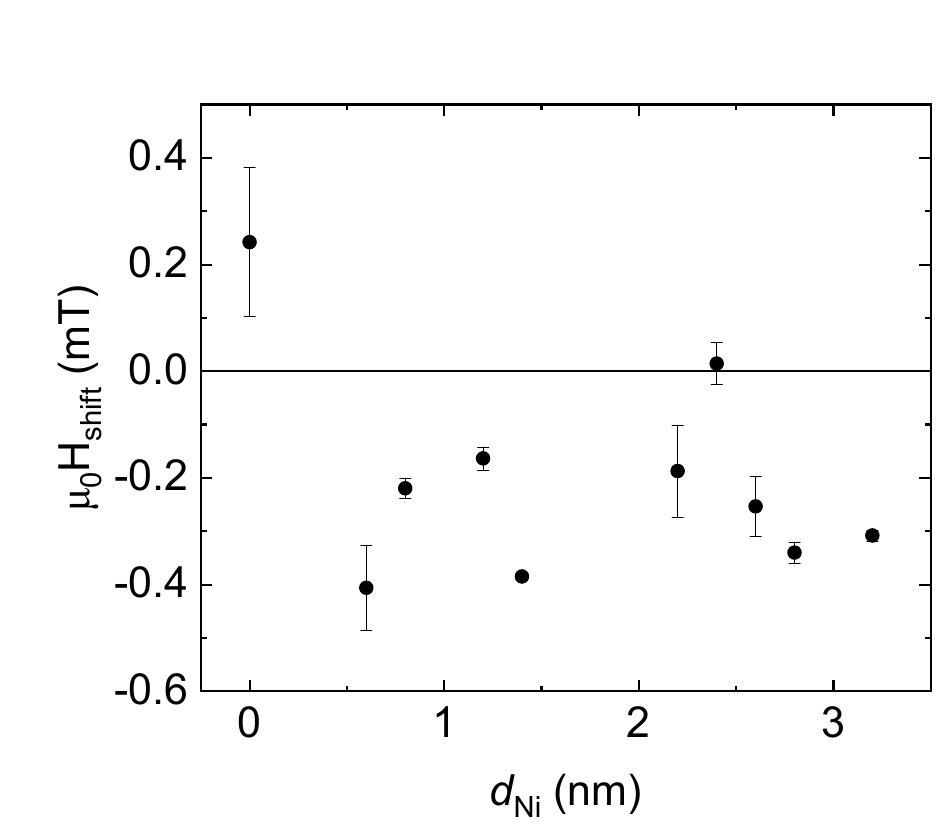}

        \caption{$\mu_0 H_\text{shift}$ for a series of preliminary samples where the DJJ is placed on a Ni ($d_\text{Ni}$)-Nb (90 nm) bilayer. Note that these data are not directly comparable to the measurements presented in the main text as the junction geometry is different and they were acquired at 4.2 K using a different experimental set up and measurement routine. Presented uncertainties represent one standard error.}
        \label{fig:prelim}
\end{figure}

In the main text we present results where the thickness of the Ni layer is fixed at 2.8 nm. This value comes from our preliminary study where DJJs are placed on a series of Ni ($d_\text{Ni}$)-Nb (90 nm) bilayers where $d_\text{Ni}$ is varied from 0.6 to 3.2 nm. The results of the preliminary study are shown in Figure \ref{fig:prelim}. From these preliminary measurements we fabricate a second set of samples, reported in the main text, with a fixed Ni thickness of 2.8 nm, as that thickness showed a significant $\mu_0 H_\text{shift}$ in the preliminary measurements. Note that these preliminary samples are not directly comparable to the samples and measurements presented in the main text. The sample geometry and measurement routines were later improved based on knowledge gained in the preliminary samples and measurements. The Ru/Al multilayer for these preliminary samples has 4 repeats. The junction measurements were acquired at 4.2 K in a He dewar using a different experimental set up to that described in the main text. Prior to the $I_c(B)$ sweep, the samples were magnetised to $\pm400$ mT followed by a warming through the sample $T_c$ to remove flux. 

\section{Meissner Screening in Low Field PNR}

In the main body of this work, we assume that the contribution of the Meissner effects to the PNR at the 1~mT applied field will be negligible to the overall magnetic response of the sample. Here, we present modelling of the data to justify such an assumption. 

We model the magnetic field expulsion as a function of depth (where \textit{z} is the distance from edge of the superconductor) in the Nb layer as determined from the London equation \cite{PenetrationNIST},

\begin{equation} \label{eq:london}
    B(z) = B_{0} \cosh{\bigg( \frac{z}{\lambda_\text{L}}} -\frac{d_{s}}{2 \lambda_\text{L}} \bigg)  \cosh{\bigg(\frac{d_{s}}{2 \lambda_\text{L}}\bigg)}^{-1},
\end{equation}

\noindent where $d_s$ and $z$ are the thickness of the superconductor and distance from the surface, respectively. Further details about these methods can be found in our previous work \cite{PhysRevMaterials.4.074801} . In this model, the external field ($B_0$) is fixed to the value measured with a Hall probe, and $\lambda_\text{L}$ is fixed to the value of 96.2~nm measured in our previous work \cite{PhysRevMaterials.4.074801}.

\begin{figure}
        \includegraphics[width=0.49\textwidth]{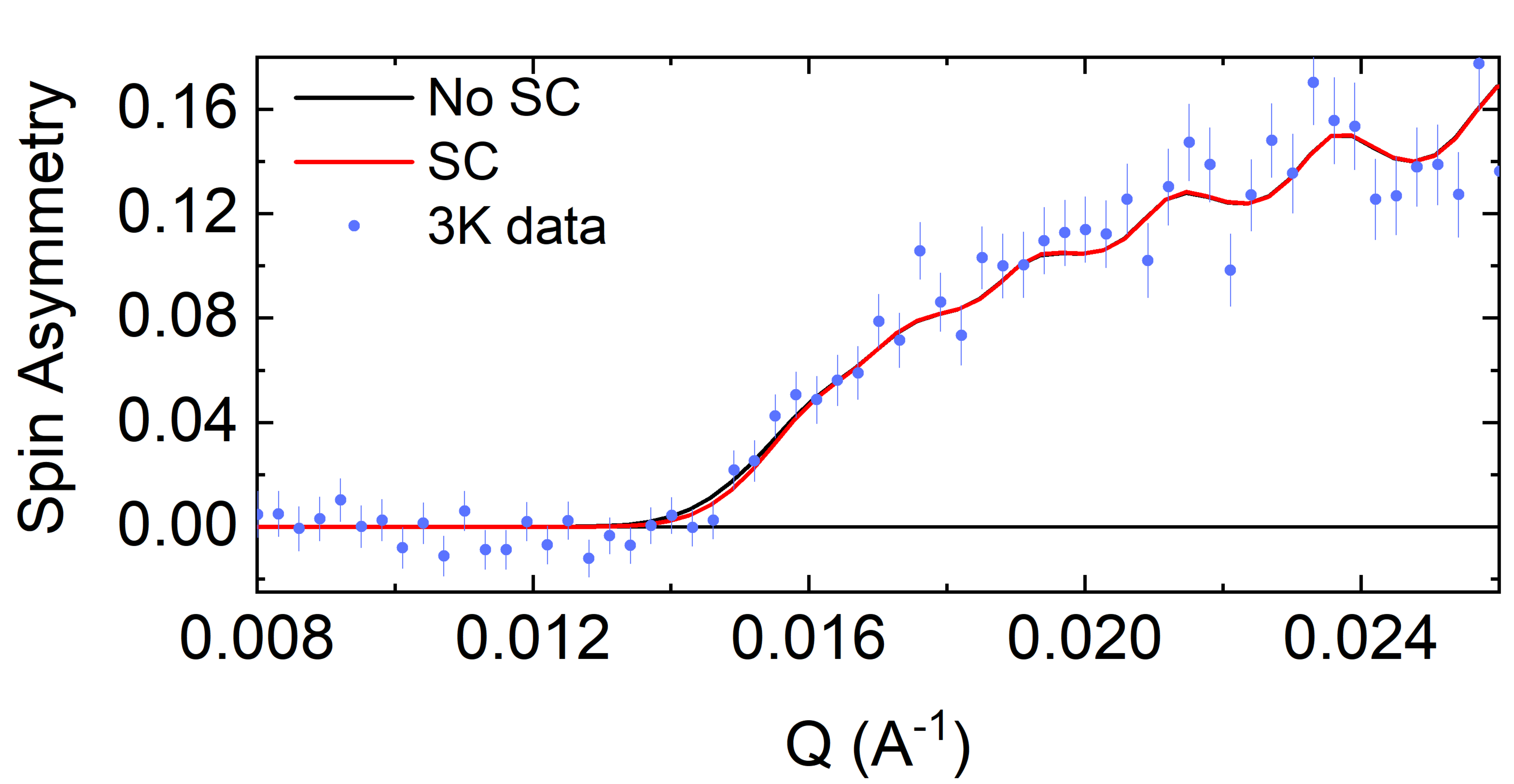}

        \caption{Low field PNR (1 mT) data on the Ni (2.8 nm)-Nb (200 nm) sample from the main body in the superconducting state at 3~K. The two fits to the data correspond to models which either do (SC) or do not (No SC) take into account a Meissner screening contribution. Presented uncertainties represent one standard error.}
        \label{fig:PNRlowfield}
\end{figure}

The results of the model are shown in Figure~\ref{fig:PNRlowfield}. The difference in the models with or without the additional contribution of Equation~\ref{eq:london} is not detectable with our experimental dataset.

\section{High Field PNR}

The main motivation of this work is to investigate the zero-field component of the EM proximity effect. The Josephson junction experiment can only probe EM proximity effect at low field due to the Fraunhofer physics of the junction. PNR on the other hand can in principle be measured at high magnetic fields. During the PNR experimental run, we also performed exploratory measurements at a high field of 0.15~T, which we report here. 

\begin{figure}
        \includegraphics[width=0.99\textwidth]{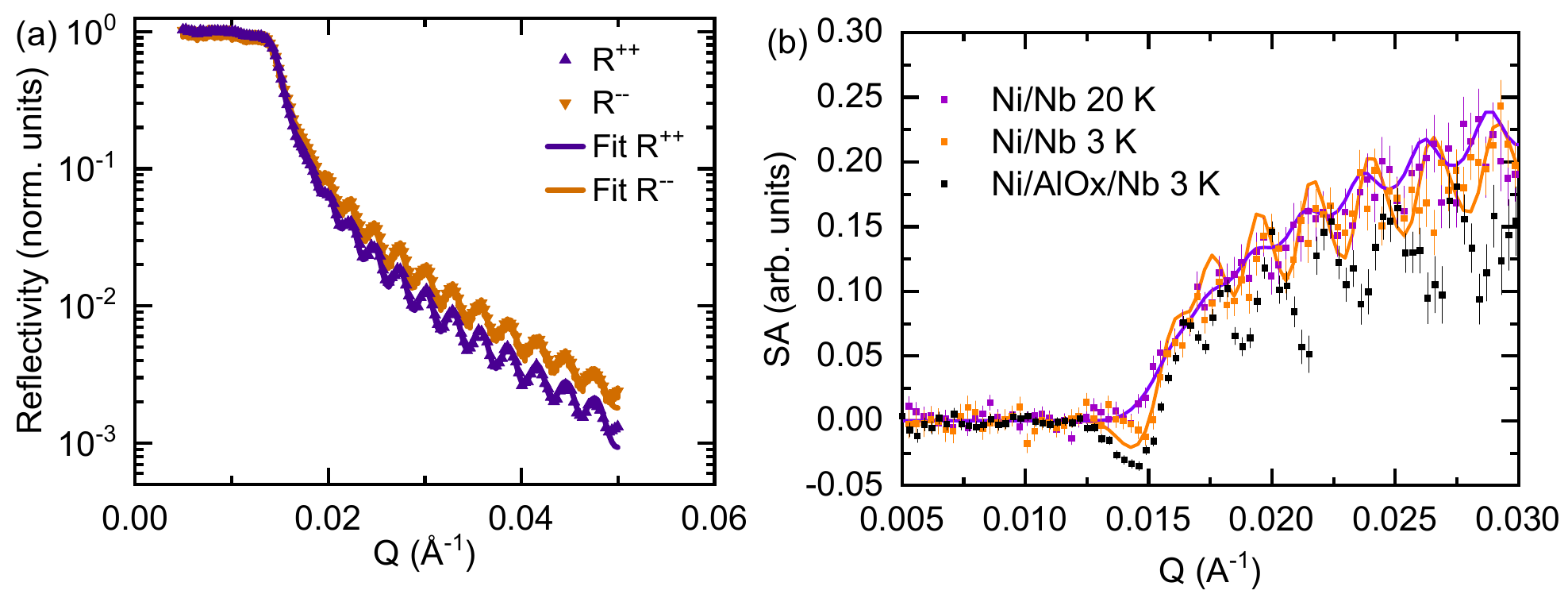}

        \caption{High field (0.15~T) polarised neutron reflectometry. (a) Non-spin-flip cross-section PNR data (points) with theoretical fits (line) at 3~K on the Ni (2.8 nm)-Nb (200 nm) sample from the main text. (b) The spin asymmetry on the Ni (2.8 nm)-Nb (200 nm) sample and a second control sample where an Al$_2$O$_3$ (2.5 nm) insulator layer is added between the Ni and Nb. Lines correspond to the best fits to the data on the Ni-Nb sample, described in the text. Presented uncertainties represent one standard error.}
        \label{fig:PNR2}
\end{figure}

Figure \ref{fig:PNR2} shows the high field (0.15~T) response of two samples. The Ni (2.8 nm)-Nb (200 nm) sample reported in the main text and a second control sample where an Al$_2$O$_3$ (2.5 nm) insulator layer is added between the Ni and Nb. In contrast to what is reported in the main text, the high field data show pronounced differences among the samples at different measurement conditions. Comparing the 20~K normal state to the 3~K superconducting state measurements of the Ni-Nb sample, upon entering the superconducting state additional features in the SA are present, particularly at Q$=0.015$\AA$^{-1}$.  

The 3 K and 20 K data for the Ni-Nb film are fit with almost all of the parameters held constant to match the fits presented at low field in the main text. The exceptions are the Ni magnetisation, as the Ni is now in the saturated state with SLD = $(1.18\pm 0.02)\times10^{-6}$~\AA$^{-2}$ (396~emu/cm$^3$) (1~emu/cm$^3=1$~kA/m) (uncertainty represents one standard error), and the Nb roughness which changes from 3~nm to 1.6~nm at 3~K. The 20~K data is fit without any superconducting component. The 3~K data is fit assuming superconductivity in the sample. The 3~K data are not well described by a model accounting for only simple Meissner screening (Equation~\ref{eq:london}) so instead we fit a profile combining linearly the expected Meissner screening and the expected profile of the EM proximity effect (Equation 1 of the main text). The field is fixed at 0.15~T, and $\lambda_\text{L}$ is fixed to the value of 96.2~nm, as above. $A_\text{EM}$, the strength of the EM proximity effect, is a free fit parameter and gives a best fit value $A_\text{EM} = -1.93$~mT. The fit (Fig. S5) is sensitive to the sign of $A_\text{EM}$, which determines whether the EM proximity effect is diamagnetic or paramagnetic in nature. If $\lambda_\text{L}$ is varied, the resulting fit gives a higher penetration depth of 102 $\pm$ 3~nm and $A_\text{EM} = -1.5 \pm 0.5$~mT (uncertainty represents one standard error). We note that this 3~K state does change slightly over extended measurement times. Also, at an applied field of 0.15~T, superconducting vortices are expected and therefore to fully describe our sample a model including vortices is necessary as it may modify the nature of the fit near the critical angle (Q$=0.015$\AA$^{-1}$) \cite{PhysRevB.80.134510, Nagy_2016, HAN2003162, PhysRevB.62.9784}, but the development of a realistic vortex model is beyond the scope of this work. Finally, attempts to fit the structure of the Ni/Al$_2$O$_3$/Nb sample have not yet been successful as the addition of oxygen during growth may have resulted in unexpected non-uniformities in the composition.

\section{Trapped flux in the superconducting magnet}

\begin{figure} []
\includegraphics[width=0.5\columnwidth]{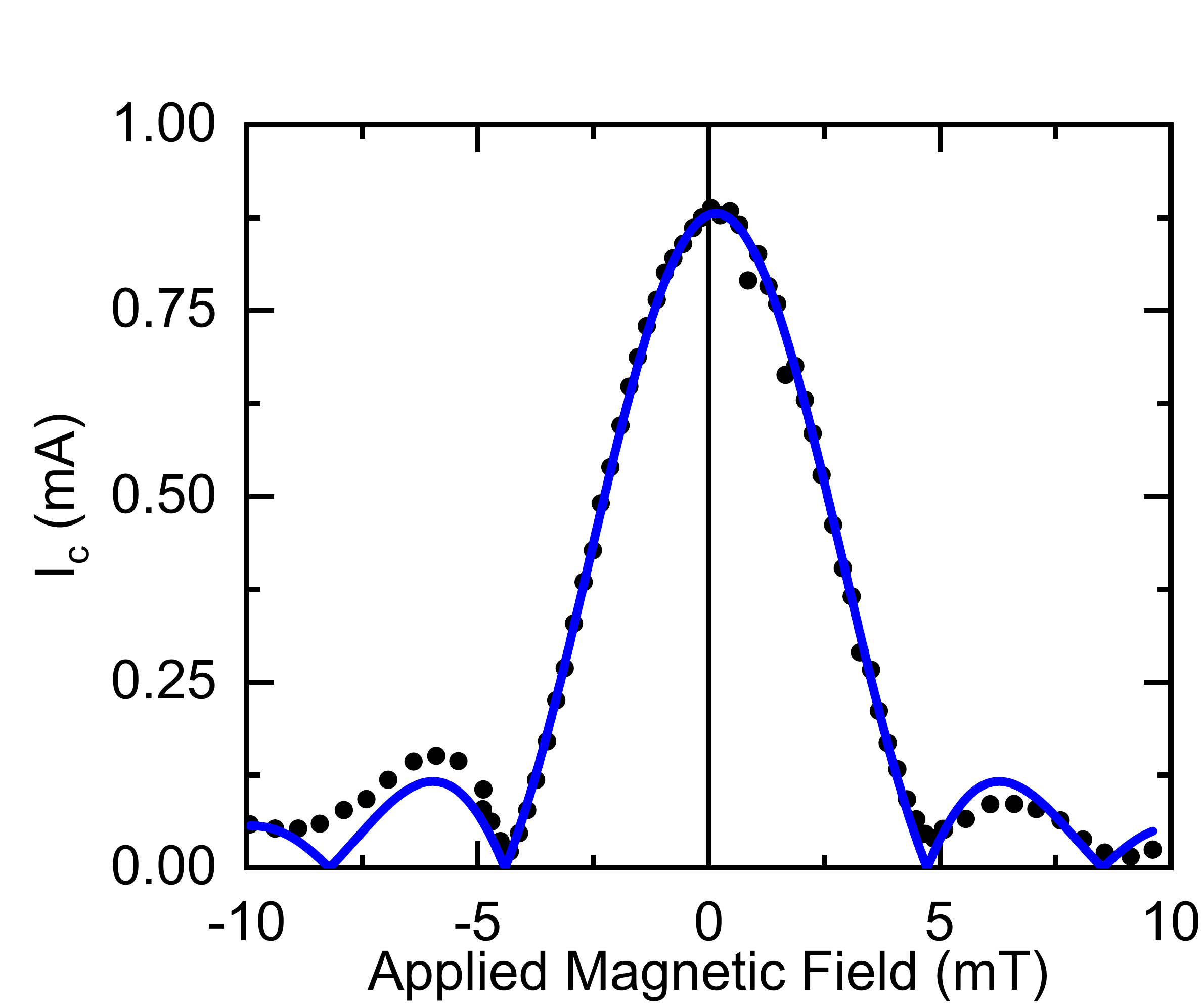} 
\caption{Product of critical Josephson current times normal-state resistance versus applied magnetic field for Josephson junctions without the Ni layer. The data are fit to Equation 2 and 3 from the main text providing a best fit for $\mu_0 H_\text{shift}=0.15$~mT, suggesting that a small positive applied field is needed to cancel the trapped flux in the superconducting coils.}
\label{SM3}
\end{figure}

Figure \ref{SM3} contains supporting data on the trapped flux in the superconducting coils used in the measurements of our samples. We have observed similar trends by measuring the field directly using a Hall probe during magnetic field sweeps when the sample space was at room temperature. The junction reported in Figure \ref{SM3} has no ferromagnetic layers. The data are fit to Equation 2 and 3 from the main text providing a best fit for $\mu_0 H_\text{shift}=0.15$~mT, suggesting that a small positive applied field is needed to cancel the trapped flux in the superconducting coils.

\section{Anomalous $I_c(B)$ Fraunhofer responses}

\begin{figure}
        \includegraphics[width=0.49\textwidth]{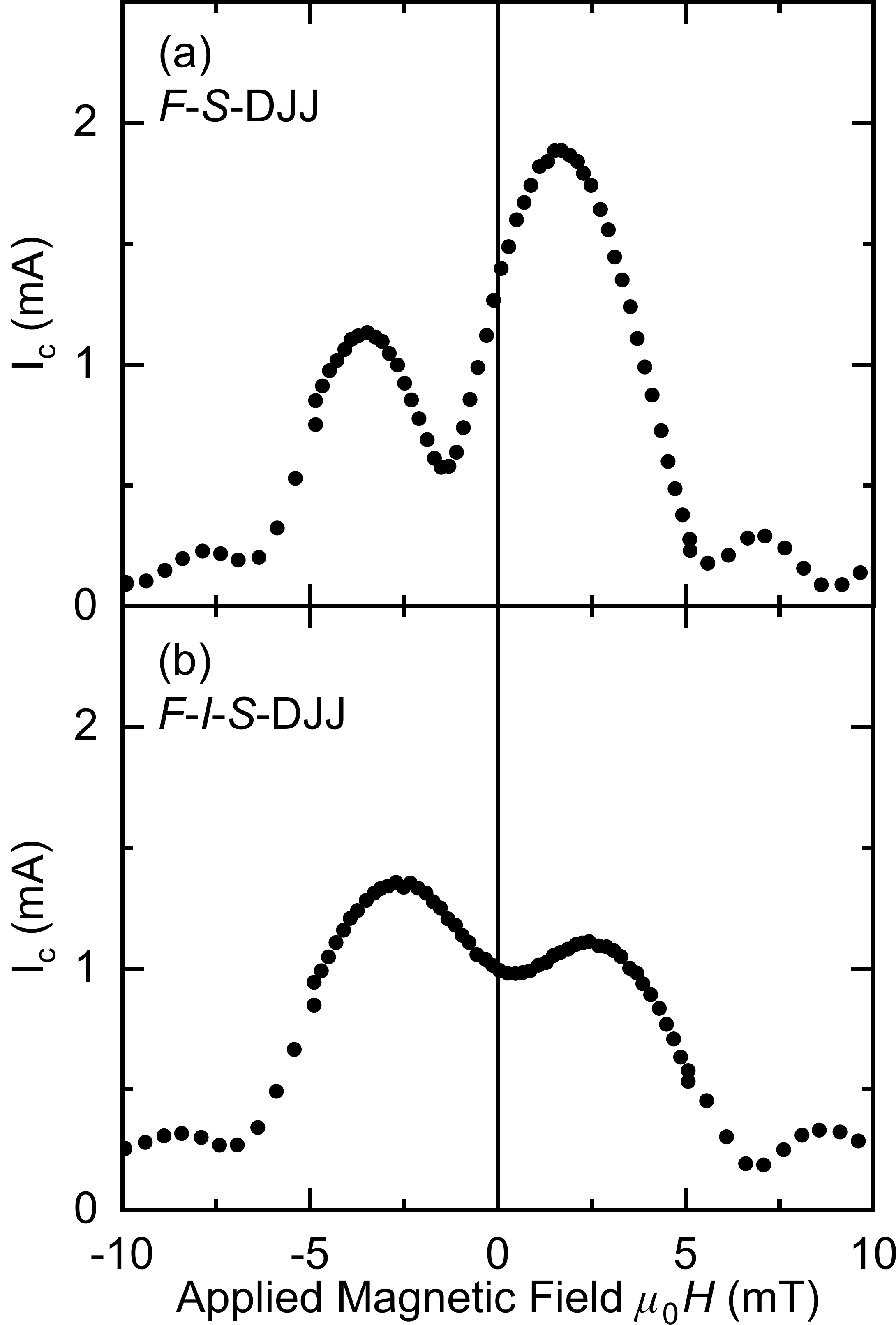}

        \caption{Selected $I_c(B)$ Fraunhofer patterns for the detection Josephson junction (DJJ) devices at 1.8~K showing the anomalous response observed in a minority of measurements. (a) The DJJ is placed on a Ni (2.8 nm)-Nb (90 nm) bilayer. (b) A control sample where an Al$_2$O$_3$ (2.5 nm) insulator layer is added between the Ni and Nb. The error in determining $I_c$ is smaller than the data points.}
        \label{fig:DJJ2}
\end{figure}

A second, anomalous, $I_c(B)$ Fraunhofer response is observed in a minority of our measurements on both the \textit{F}-\textit{S} and \textit{F}-\textit{I}-\textit{S} control samples, shown in Figure \ref{fig:DJJ2}. The devices of \ref{fig:DJJ2} are the same devices as reported in the main body of this paper and shown in Figure 3. This second type of response is characterised as having a non-uniform $I_c(B)$, which is not described well by the expected Airy function (Equations 2 and 3). In the examples shown in Figure \ref{fig:DJJ2}, the peak $I_c(B)$ is also reduced from the measurements in Figure 3.

Since the anomalous $I_c(B)$ response is observed in both the \textit{F}-\textit{S} and \textit{F}-\textit{I}-\textit{S} control samples, we can rule out an electronic proximity effect origin in this case also. To investigate the origin, we performed further measurements varying the initialisation routine. We report that the anomalous $I_c(B)$ response can be consistently observed by demagnetising the sample prior to measurement. The demagnetising routine generates a multidomain state in the Ni layer. We therefore conclude that the anomalous $I_c(B)$ response is a result of stray fields when a domain wall in the Ni layer forms directly below the DJJ. Since the magnetic remanence of the Ni layer is $\approx70$\% (Figure 1), even after the saturating initialisation routine there is still a chance that a domain wall is generated below the DJJ.

These observations are somewhat comparable to existing literature on \textit{S}-\textit{F}-\textit{S} Josephson junctions with multidomain \textit{F} barriers, where the stray fields distort the Fraunhofer pattern \cite{Bourgeois, PhysRevB.79.094523, PhysRevB.80.020506}. Such an observation is somewhat surprising here, since the geometry of our devices is quite different to the previous studies. 

\begin{figure}
        \includegraphics[width=0.49\textwidth]{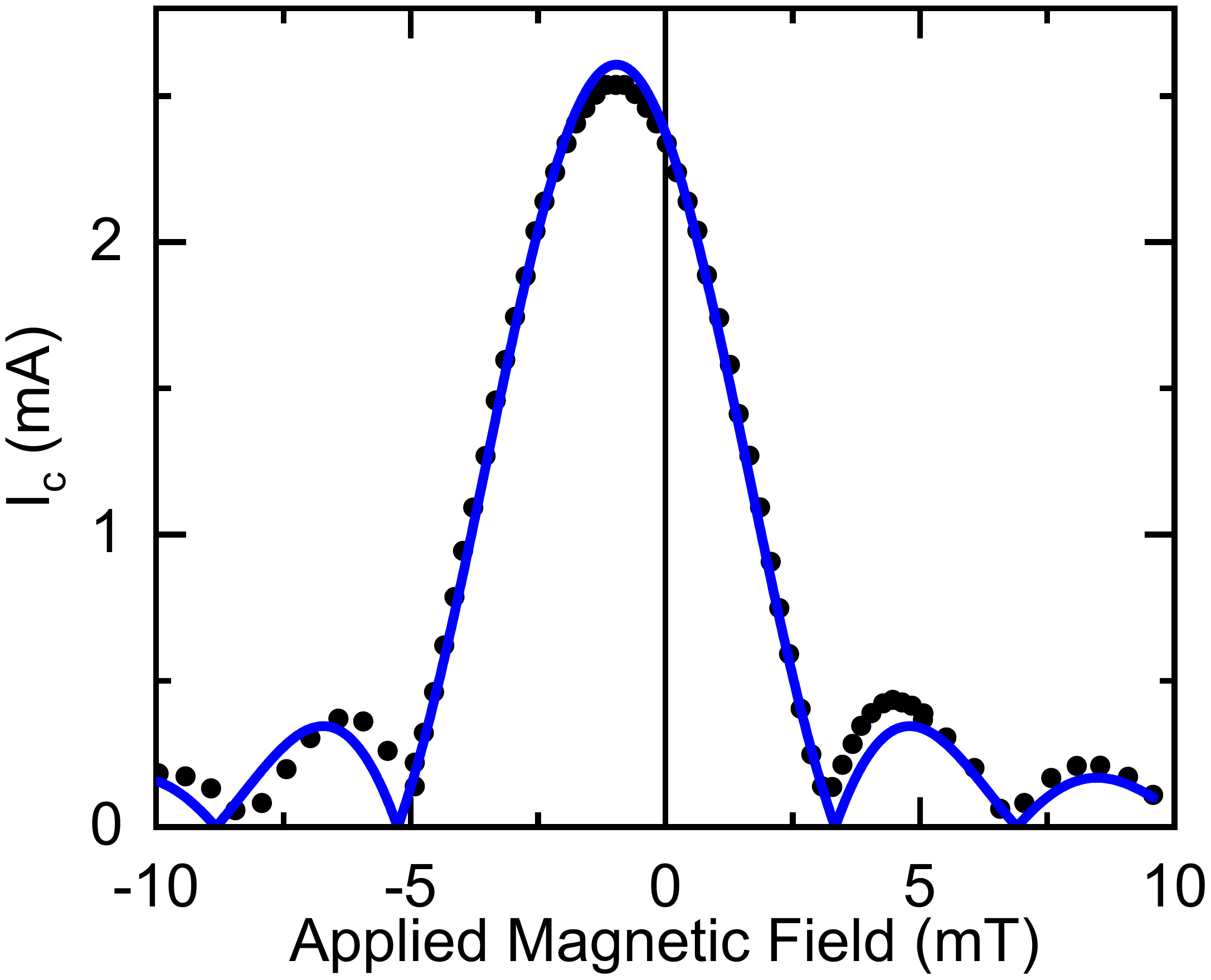}

        \caption{$I_c(B)$ Fraunhofer pattern at 1.8~K observed in a minority of measurements on one of the three measured Ni (2.8 nm)-Nb (90 nm) bilayer samples. This particular sample showed three types of $I_c(B)$ response, the response reported in the main text, the anomalous $I_c(B)$ response of Figure \ref{fig:DJJ2}, and a third behaviour presented here. This measurement showed a significant $\mu_0 H_\text{shift}=0.96\pm0.01$~mT (uncertainty represents one standard error). This $I_c(B)$ response is not reproduced in the other two samples measured in this study and is hence not included in main text or calculations therein. The error in determining $I_c$ is smaller than the data points.}
        \label{fig:DJJ3}
\end{figure}

A third $I_c(B)$ response is observed in one of the three \textit{F}-\textit{S} samples in a minority of measurements, shown in Figure \ref{fig:DJJ3}. This particular sample showed all three types of $I_c(B)$ response; the response reported in the main text, the anomalous $I_c(B)$ response of Figure \ref{fig:DJJ2}, and a third behaviour presented here. This behaviour showed an $I_c(B)$ response with a well defined Fraunhofer pattern and a significant $\mu_0 H_\text{shift}=0.96\pm0.01$~mT averaged over five measurements (uncertainty represents one standard error). This behaviour is not reproduced in the other two \textit{F}-\textit{S} samples measured in this study, and so these five $I_c(B)$ responses are not used in the calculation of average $\overline{H}_\text{shift}$ values reported in the main text. This behaviour was also not observed in any of the three \textit{F}-\textit{I}-\textit{S} control samples.

\section{Temperature dependent transport}

\begin{figure}
        \includegraphics[width=0.49\textwidth]{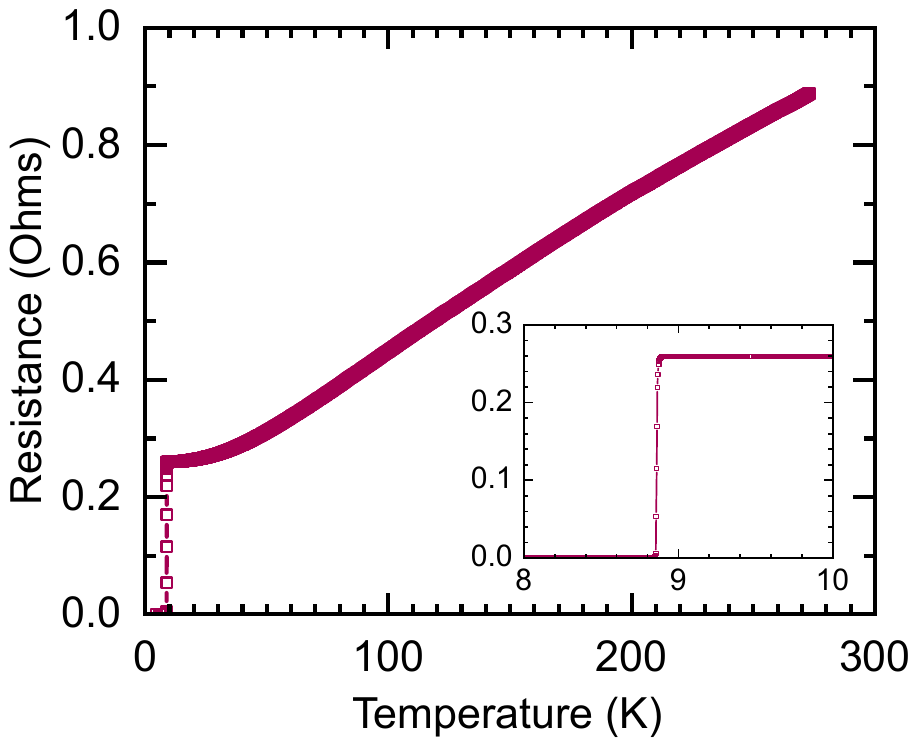}

        \caption{Electrical transport of the Ni (2.8 nm)-Nb (200 nm) bilayer sample. The inset shows the superconducting transition.}
        \label{fig:RvT}
\end{figure}

The resistance of the sheet film Ni (2.8 nm)-Nb (200 nm) bilayer sample is measured as a function of temperature. Four electrical contacts are made in a line in the center of the sample by wire bonding for traditional four-point probe transport with a measurement current of 1~mA. In the main text, the normalised resistance close to the transition temperature is presented. Figure \ref{fig:RvT} shows the full range of the transport data. The residual resistivity ratio of the sample is determined to be 3.4 from these measurements. 

\section{Magnetic hysteresis loops}

\begin{figure}
        \includegraphics[width=0.98\textwidth]{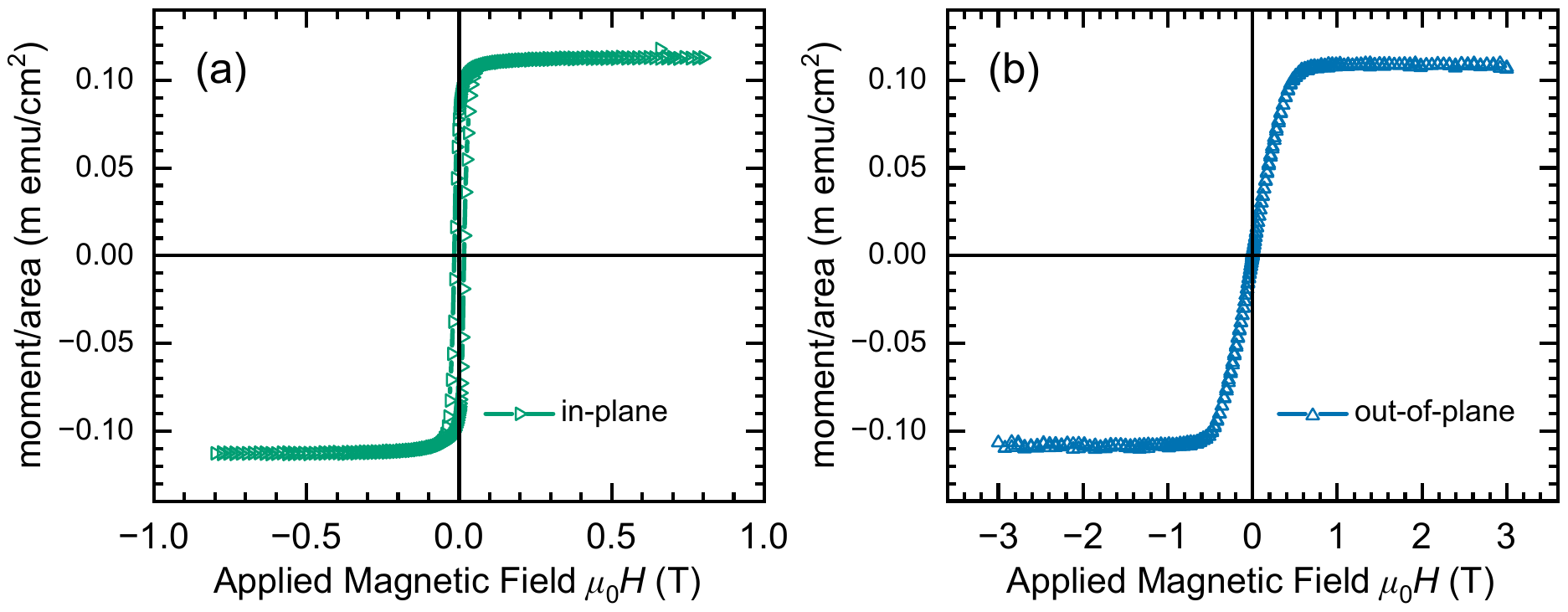}

        \caption{Magnetic characterisation of the Si(sub)-Ni(2.8)-Nb(200) sheet film sample. Magnetic hysteresis loops acquired at a temperature of 10 K with the applied field oriented (a) in-plane and (b) out-of-plane. The diamagnetic contribution from the substrate has been subtracted. Moment/area is calculated from the measured total moment and the measured area of the sample cutting.}
        \label{fig:MagFull}
\end{figure}

In the main text we present in- and out-of-plane magnetic hysteresis loops showing the applied fields close to the switching fields. Here, we plot the hysteresis loops over the full acquired range of fields. Figure \ref{fig:MagFull} (a) shows the in-plane field orientation and (b) the out-of-plane applied field orientation of the Si(sub)-Ni(2.8)-Nb(200) sample at 10~K. The diamagnetic signal from the substrate has been subtracted. The moment/area is calculated from the measured total moment of the sample and the measured area of the cutting used.

\bibliography{EMProx}